%% file: 0_main.tex
\documentclass[sigconf]{acmart}
\usepackage{siunitx}
\usepackage{sourcecodepro}
\usepackage[T1]{fontenc}
\usepackage{listings}
\setcopyright{none}
\acmISBN{}
\acmConference[1st International Workshop on Low Carbon Computing (LOCO 2024)]{}{Dec 03, 2024}{Glasgow, Scotland, UK}

\AtBeginDocument{\DeclareSIUnit{\kWh}{kWh}}

\begin{document}

\title{Emission Impossible: privacy-preserving carbon emissions claims}

\author{Jessica Man}
\affiliation{%
  \institution{Department of Computer Science \& Technology\city{Cambridge}\country{UK}}
}
\email{psjm3@cam.ac.uk}

\author{Sadiq Jaffer}
\affiliation{%
  \institution{Department of Computer Science \& Technology\city{Cambridge}\country{UK}}
}
\email{sj514@cam.ac.uk}

\author{Patrick Ferris}
\affiliation{%
  \institution{Department of Computer Science \& Technology\city{Cambridge}\country{UK}}
}
\email{pf341@cam.ac.uk}

\author{Martin Kleppmann}
\affiliation{%
  \institution{Department of Computer Science \& Technology\city{Cambridge}\country{UK}}
}
\email{martin.kleppmann@cst.cam.ac.uk}

\author{Anil Madhavapeddy}
\affiliation{%
  \institution{Department of Computer Science \& Technology\city{Cambridge}\country{UK}}
}
\email{avsm2@cl.cam.ac.uk}

\renewcommand{\shortauthors}{Man et al.}
\begin{abstract}
\input{1_abstract}
\end{abstract}


\maketitle

\section{Introduction}
\input{2_introduction}

\section{Background}
\input{3_background}

\section{Related Work}
\input{4_related_work}

\section{A ZKP Emissions Disclosure Scheme} \label{sec:methodology}
\input{5_methodology}

\section{Conclusion}
\input{6_discussion}

\bibliographystyle{ACM-Reference-Format}
\bibliography{0_main}

\end{document}

%% file: 1_abstract.tex
Information and Communication Technologies (ICT) have a significant climate impact, and data centres account for a large proportion of the carbon emissions from ICT. To achieve sustainability goals, it is important that all parties involved in ICT supply chains can track and share accurate carbon emissions data with their customers, investors, and the authorities. However, businesses have strong incentives to make their numbers look good, whilst less so to publish their accounting methods along with all the input data, due to the risk of revealing sensitive information. It would be uneconomical to use a trusted third party to verify the data for every report for each party in the chain. As a result, carbon emissions reporting in supply chains currently relies on unverified data. This paper proposes a methodology that applies cryptography and zero-knowledge proofs for carbon emissions claims that can be subsequently verified without the knowledge of the private input data. The proposed system is based on a zero-knowledge Succinct Non-interactive ARguments of Knowledge (zk-SNARK) protocol, which enables verifiable emissions reporting mechanisms across a chain of energy suppliers, cloud data centres, cloud services providers, and customers, without any company needing to disclose commercially sensitive information. This allows customers of cloud services to accurately account for the emissions generated by their activities, improving data quality for their own regulatory reporting. Cloud services providers would also be held accountable for producing accurate carbon emissions data.

%% file: 2_introduction.tex
Regulatory requirements and sustainability initiatives mean that companies are increasingly having to report their carbon emissions. Looking at ICT companies in particular, customers of online services, who are also obligated to report their emissions data or who might want to take carbon emissions into account when deciding which service to use, are currently hindered by a lack of reliable emissions data that are comparable across services. Calculating accurate carbon emissions across a cloud computing pipeline involves a number of stakeholders, none of whom are incentivised to accurately report their emissions for competitive reasons. In this paper, we explore mechanisms to support verifiable and confidentiality-preserving emissions reporting across a chain of energy suppliers, cloud data centres, virtual machine hosting service providers and cloud services providers, which are ultimately passed through to their customers. We believe that adding verifiable and composable emissions transparency to cloud computing architectures enables providers to compete on the basis of sustainability, resulting in demand-side pressure on cloud services to shift to renewable energy sources \cite{calderon2020analysis}.

Our technique centres around zero-knowledge proofs (ZKPs) \cite{goldwasser2019knowledge}. When applying ZKPs to the issue of untrusted carbon emissions claims, a stakeholder in a supply chain proves to a verifier (who can be anyone, such as a customer, investor, or regulator) that the emissions calculations were performed accurately, without revealing commercially sensitive data about their business operations. The verifier decides whether the claim can be accepted using only public knowledge and the cryptographic proof provided by the stakeholder. The proposed system applies zero-knowledge Succinct Non-interactive ARguments of Knowledge (zk-SNARK) as the protocol to allow proofs to be generated and subsequently verified without the requirement of disclosing all of the input data.

In this paper, we present our argument on why conflicting incentives around carbon emissions reporting make existing systems unlikely to succeed (\S\ref{sec:conflicting_incentives}) and our contribution of applying ZKPs to allow more accurate reporting of carbon claims without compromising sensitive information (\S\ref{sec:methodology}).

%% file: 3_background.tex
\subsection{Carbon Emissions Reporting}
In some countries, large companies must disclose their carbon emissions to comply with regulations. For example, the UK's Streamlined Energy and Carbon Reporting (SECR) regulations require all UK quoted companies and large limited liability partnerships to report on their global energy use in addition to greenhouse gas emissions \cite{secr}. The EU Corporate Sustainability Reporting Directive (CSRD) came into force on 5 January 2023, which requires large companies and listed SMEs to report on sustainability, highlighting the urgent need for the disclosure of `relevant, comparable and reliable sustainability information' and the significant increase in demand for sustainability information \cite{csrd-official}. Other than for regulatory reasons, companies publicising their plans towards net-zero could have a positive impact on their businesses. On the one hand, transparency on climate actions taken as part of the manufacturing and distribution process behind commercial products can directly influence consumer behaviours \cite{calderon2020analysis}. On the other hand, if companies have to disclose their confidential business data as part of the reporting, this could benefit competitors.

Pressure from the media, investors, and customers could also have an effect on businesses. Amazon was given an `F' grade (meaning no response) by CDP\footnote{A non-profit carbon disclosure company, formally known as the Carbon Disclosure Project} until 2023 when they first submitted their report \cite{reveal_amazon_undercounts}. Shareholders of the company continue to request Amazon to provide additional information on climate-related impacts \cite{amazon_stakeholders_request,amazon_shareholders_meeting_2024}.

\subsection{Conflicting Incentives Behind CO\textsubscript{2e} Claims} \label{sec:conflicting_incentives}
An estimated 1.8\% to 3.9\% of global carbon emissions are attributable to Information and Communication Technology (ICT) \cite{acm_ict_on_ghg}.  Governments, investors and customers are therefore paying close attention to how cloud computing operators work towards net zero \cite{azar2021big, gs_datacenter, gelenbe2015impact}. However, companies closely guard the metrics behind the computational resources used to provide particular services, as disclosing them would undermine their pricing and business model. Businesses have strong incentives to make only positive claims, which could involve hiding data or publishing misleading results with dubious evidence, a problem termed `greenwashing'~\cite{greenwashing}. Therefore, it is difficult for an outsider to know whether a company's emissions claims are true.

There are now tightened regulations to tackle greenwashing in some countries, for example, the EU has proposed the `Green Claims Directive' \cite{eu_gcd} to prevent companies making claims without providing clear evidence. The verification, however, relies on accredited verifiers. The UK's CMA has also developed the `Green Claims Code' to combat greenwashing\footnote{https://greenclaims.campaign.gov.uk}, which focuses on a set of core principles based on existing consumer law, to protect businesses and consumers from misleading environmental claims. However, measurements of carbon emissions are continuous and frequent, and with input data varying at each measurement, it is not practical to have an independent auditor to verify every single claim for a supply chain with multiple companies involved.

Consider three of the biggest data centre providers: Amazon, Google and Microsoft. They have all reported their decarbonisation goals publicly, but have also been accused of using creative accounting to hide facts about their carbon emissions \cite{mitreview_bigtech_carbon_claim}. Both Microsoft and Google admitted that their carbon emissions had increased in recent years, despite their climate commitments \cite{theregister_google_emissions_up,theregister_microsoft_emissions_up}, and Amazon's self-reporting did not include emissions data for products sold by third-party vendors \cite{reveal_amazon_undercounts}. The major risks involved in data centre emissions reporting are:

\begin{enumerate}
    \item \textbf{Privacy and trade secrets concerns.} Reports on carbon emissions typically show only aggregated data at high levels. Validation of claims often requires details of carbon-emitting activities, which cannot be publicly disclosed because they are trade secrets of both suppliers and customers. 
    \item \textbf{Untrustworthy claims.} Companies make misleading claims based on dubious accounting methodologies to make it look like they are more environmentally friendly than they actually are \cite{greenwashing}. Yang et al.~studied greenwashing behaviours and impact and found that greenwashing is often linked with scandals that occur at the supply chain level \cite{yang2020greenwashing}.
    \item \textbf{Missing claims.} Companies can choose not to disclose anything or report claims that omit some of their emissions-generating activities. Amazon's undercounting in their carbon footprint reports is a good example \cite{reveal_amazon_undercounts}.
\end{enumerate}

%\subsection{Mitigations}
We can mitigate the first risk to protect businesses by ensuring that the verification method does not leak secrets. For the second risk, we can use verified data to provide trustworthy claims. Companies can provide proofs that their claims on the carbon emissions report are all true, and the proofs can be checked out by their customers, investors or auditors. ZKPs (\S\ref{sec:related_work_zkp}) can be used for these mitigations. To ensure figures are comparable across different companies, the calculation methodology can be standardised, as discussed in \S\ref{sec:methodology}. For the third risk, we cannot validate missing data. We can, however, bind carbon accounting to financial accounting to make it more difficult to cheat, as discussed in \S\ref{sec:caveats_and_future_work}.  Finally, we can analyse and benchmark the verified data across companies and look for discrepancies and anomalies through manual audits. Manual audits are carried out infrequently, typically annually, and it is a lengthy process. Therefore, while we cannot totally eliminate the need for manual audits, ZKPs provide a complementary process that can be automated for much faster and more frequent verifications.

%% file: 4_related_work.tex
\subsection{Carbon Accounting and Reporting} \label{sec:relate_work_accounting}
The most commonly used approach to calculate carbon emissions is to follow the Greenhouse Gas (GHG) Protocol \cite{ghg_reporting_standards}. According to the GHG Protocol Corporate Standard, emissions are categorised in three scopes: Scope 1 emissions occur from sources that are owned or controlled by the company; Scope 2 emissions are generated from purchased electricity or heat consumed by the company; Scope 3 emissions are a consequence of the activities of the company, but occur from sources not owned or controlled by the company, primarily by its direct or indirect suppliers. In most companies, Scope 3 accounts for the majority of emissions by far \cite{pact}, but is also the most complex to compute, given that the calculation of emissions requires data from the entire supply chain. For those companies that report their carbon emissions, most of them currently report on Scope 1 carbon emission, less so on Scope 2 and very little on Scope 3 \cite{pact}.

When an organisation migrates their on-premises computing resources and IT workload to the cloud, emissions under scope 1 or scope 2 move to scope 3. Given that scope 3 emissions reporting is voluntary and data centres' emissions are aggregated into the global reporting by large cloud providers, cloud service customers' emissions become hidden \cite{mytton2020hiding}. Customers therefore rely on cloud providers to tell them their share of emissions arising from the data centre for the hosted services they use, but the lack of transparency has made this a big challenge for the customers. If the information is not provided by the cloud providers, customers have to estimate using aggregated global data and other methods to obtain information, such as through Freedom of Information requests, or make in-house measurements to produce the metrics themselves \cite{etsy_cloud_jewels, aws_track_power_hard}.

In recent years, the big cloud providers have started to provide their customers with more detailed reports about the emissions of the computing resources they use. For example, Microsoft offers an `Emissions Impact Dashboard' tool to their customers \cite{microsoft_dashboard}, with the methodology verified by a third-party company \cite{microsoft_thirdparty_verified}. Google also offers a similar tool to their cloud customers and has published their carbon accounting methods \cite{schneider2024carbon}, and the methodology has also been reviewed by a third party company \cite{google_thirdparty_verified}. Neither of these tools provides a way for customers to independently verify the reported data. The review process by the cloud providers' appointed trusted third-parties does not validate the reported data for all customers (only a sample during the review period).

WBCSD\footnote{https://www.wbcsd.org} (The World Business Council for Sustainable Development) was formed in 1995 and provides a platform for businesses around the globe to respond to sustainability challenges. WBCSD has over 200 leading business members globally across multiple industries, working together to create standards, policies, and best practices that drive the sustainability agenda. They published the `Partnership for Carbon Transparency' (PACT) methodology to provide guidance on carbon accounting and exchange of emissions data, and highlighted that assurance and verification are key in ensuring the credibility and reliability of exchanged data \cite{pact}. Their methodology involves using a third-party provider for the verification; our proposal could complement that approach by providing an automatic approach to protecting confidential data with a high level of confidence.

Heiss et al.~\cite{heiss2024VerifiableCarbonAccounting} propose using zero-knowledge proofs to provide verifiable data in carbon emissions accounting, while protecting confidential business data. Their method applies to product carbon accounting, and they used the automotive industry as a practical use case to demonstrate at a high level how the approach would work. Their design extends the `Digital monitoring, reporting and verification (D-MRV)' systems, which rely on blockchains to produce authenticity and integrity proofs.

% Look up definition and correct what argument of knowledge means, it's not proof of knowledge.
\subsection{Zero-Knowledge Proofs (ZKPs)} \label{sec:related_work_zkp}
We propose using a zero-knowledge Succinct Non-interactive ARguments of Knowledge (zk-SNARK) protocol \cite{ben2014succinct}. zk-SNARK protocols allow a prover to convince a verifier that the prover knows values (a \emph{witness}) that satisfy a given set of equations (a \emph{circuit}) in \emph{zero knowledge}, i.e.\ without revealing any information about the witness \cite{ben2014succinct,bitansky2012extractable,7546506}. zk-SNARKs are \emph{succinct}: informally, this means that the proof is small compared to the witness and fast to verify \cite{zkjargon}. The protocol is \emph{non-interactive}, which is similar to digital signatures in that the prover only needs to send a single message to the verifier (the proof). The \emph{ARK} part of zk-SNARK means that the prover can convince the verifier that it knows the witness. This property is formalised as \emph{knowledge soundness}, which means that a computationally bounded prover cannot generate a proof of a false statement, or a statement for which it does not have a witness, except with negligible probability \cite{zkjargon}.

zk-SNARKs are typically applied in systems for managing digital assets, where transactions are executed without sensitive information such as the origin or amount of the transaction being revealed \cite{huang2022zkchain, sasson2014zerocash, hopwood2016zcash}. Zcash \cite{sasson2014zerocash, hopwood2016zcash} was one of the first widespread applications of zk-SNARKs, allowing transactions to be fully encrypted on a blockchain and still be verifiable. zk-SNARK has also been applied to protect medical data. Luong and Park \cite{luong2022privacy} proposed a blockchain-based system with IoT devices, allowing data sharing without leaking patient records.

Blockchains incur large infrastructure costs due to their requirement to establish consensus on a global ledger of transactions in a trustless setting. However, our emissions reporting use case does not require such a global ledger, since emissions data can be exchanged directly between suppliers and customers. We can therefore avoid the costs of blockchains in our system.

%% file: 5_methodology.tex
The core challenge in applying ZKPs to carbon emissions reporting is ensuring that the data in the ZKP is an accurate reflection of reality. ZKPs can only check that the prover knows some input values, but those values could be made up. We approach this problem by identifying ways of cryptographically proving the correctness of every input to the emissions calculation, for example by assuming that it is signed by a trusted authority (whose public key becomes a root of trust), or by requiring the originator of an input value to also provide a SNARK proof of its validity, which can be checked recursively.

\subsection{Data Centre Use Case} 
To illustrate, consider a scenario where a user wants to compare the carbon emissions of AI chatbots such as ChatGPT (OpenAI), Gemini (Google) and Claude (Anthropic). To ensure a fair comparison, they must be computed according to the same methodology, and each emissions claim must be verified end-to-end.

\begin{figure*}
    \centering
    \includegraphics[width=\linewidth]{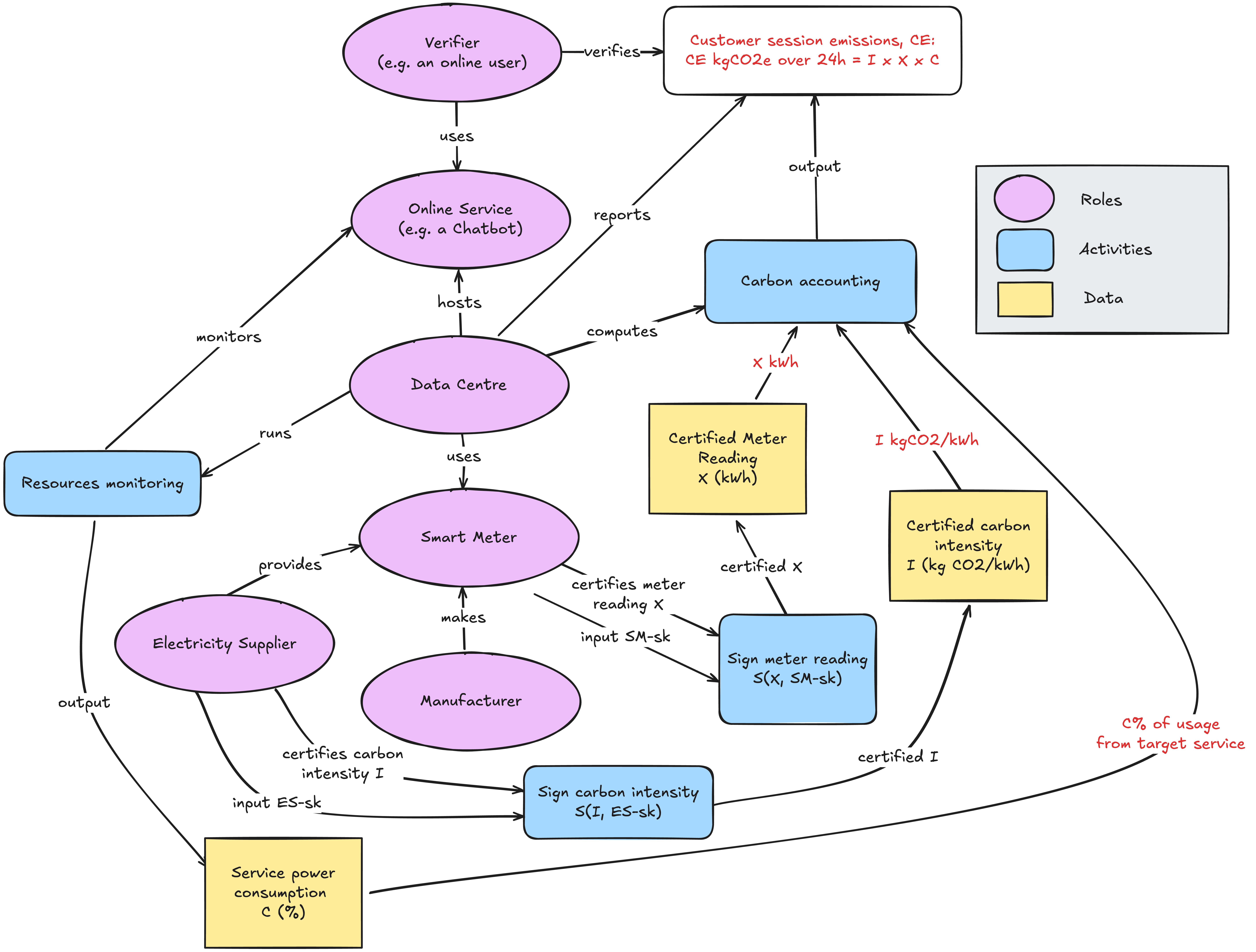}
    \caption{A simplified carbon accounting flow overview}
    \label{fig:simplified_carbon_accounting}
\end{figure*}

We can model a simplified scenario for data centre emissions reporting as illustrated in Fig.~\ref{fig:simplified_carbon_accounting}. In this scenario, we only consider electricity as the source of emissions. In reality, there are other sources such as embodied emissions from hardware manufacturing, for example, but electricity supply is currently the dominant factor in cloud computing emissions \cite{schneider2024carbon}.

To convert energy use (measured in $\mathrm{kWh}$) into emissions (measured in $\mathrm{kgCO}_{2e}$) we also need to know the \emph{carbon intensity} (measured in $\mathrm{kgCO}_{2e}/\mathrm{kWh}$). The carbon intensity of the electricity grid varies by time and place (for example, it is lower when lots of renewables are available), and it is affected by the datacenter operator's commercial arrangements (e.g.\ commitments to buy a certain amount of energy from a low-carbon supplier, or the operator may even generate their own power). For simplicity, we assume we can use grid average carbon intensity data provided by the data centre's electricity supplier.

As a further simplification, we focus on the emissions of services that are the direct result of a particular customer's usage of data centre resources. We leave for future work the question of how to deal with large fixed-cost emissions, such as those from training the AI models, which need to be amortised over the useful lifetime of the model.

We can thus define the carbon emissions calculation for a user of cloud computing services as follows:
% I is too close to the number '1', maybe change it? break the formula down in a few steps - without measurement units first with full name of each, then add them 
% \begin{equation*}
%     CE\mathrm{\hspace{0.1cm}kgCO_{2e}} = I\mathrm{\hspace{0.1cm}kgCO_{2e}}/\mathrm{kWh} \cdot X\mathrm{\hspace{0.1cm}kWh} \cdot C\%
% \end{equation*}
\begin{equation*}
    \mathit{CustomerEmissions} = \mathit{Intensity}\cdot\mathit{TotalEnergy}\cdot\mathit{Share}
\end{equation*}
or as presented using one-letter symbols on Fig.~\ref{fig:simplified_carbon_accounting}:
\begin{equation*}
    CE = I \cdot X \cdot C
\end{equation*}

The data centre operator should be able to send each of their customers the value $\mathit{CustomerEmissions}$, the carbon emissions resulting from that customer's resource usage, without that customer learning anything about other customers' emissions or the data centre's total emissions. Let $\mathit{Share}$ be the fraction of the data centre's total energy consumption allocated to a particular customer over the reporting period. The carbon intensity could be public knowledge if a grid average is used, but it might be sensitive if it reflects commercial arrangements with electricity suppliers. The data centre's total energy consumption $\mathit{TotalEnergy}$ and a particular customer's fraction of it $\mathit{Share}$ are commercially sensitive because they reflect the data centre operator's business volume and profitability.

All the figures are averages over the same fixed period of time, e.g. over 24 hours. 

\subsection{Verifiable Inputs}
% Make it explicit that the workflow relies on security of public-key-based encryption and validation chain.
To make it more difficult to cheat on the electricity usage figures, we can assume that the data centre operator uses smart meters with a secure hardware element that signs meter readings with a private key configured by the meter manufacturer. To ensure that the public key used to verify the meter reading signatures represents a genuine meter, we need the smart meter's public key to be signed by the meter manufacturer. The meter manufacturer's public key can in turn be signed by a trusted Certificate Authority (CA) that has checked that the public key belongs to a reputable manufacturer of smart meters. Similarly to a TLS certificate chain, this approach can be used to verify signed meter readings and prevent tampering and forgery. We have to assume that the smart meter is correctly installed and not bypassed, but since the meter readings are also needed for billing, it is in the electricity supplier's interest to check the correct installation of the meter.

\begin{figure*}
    \centering
    \includegraphics[width=0.5\linewidth]{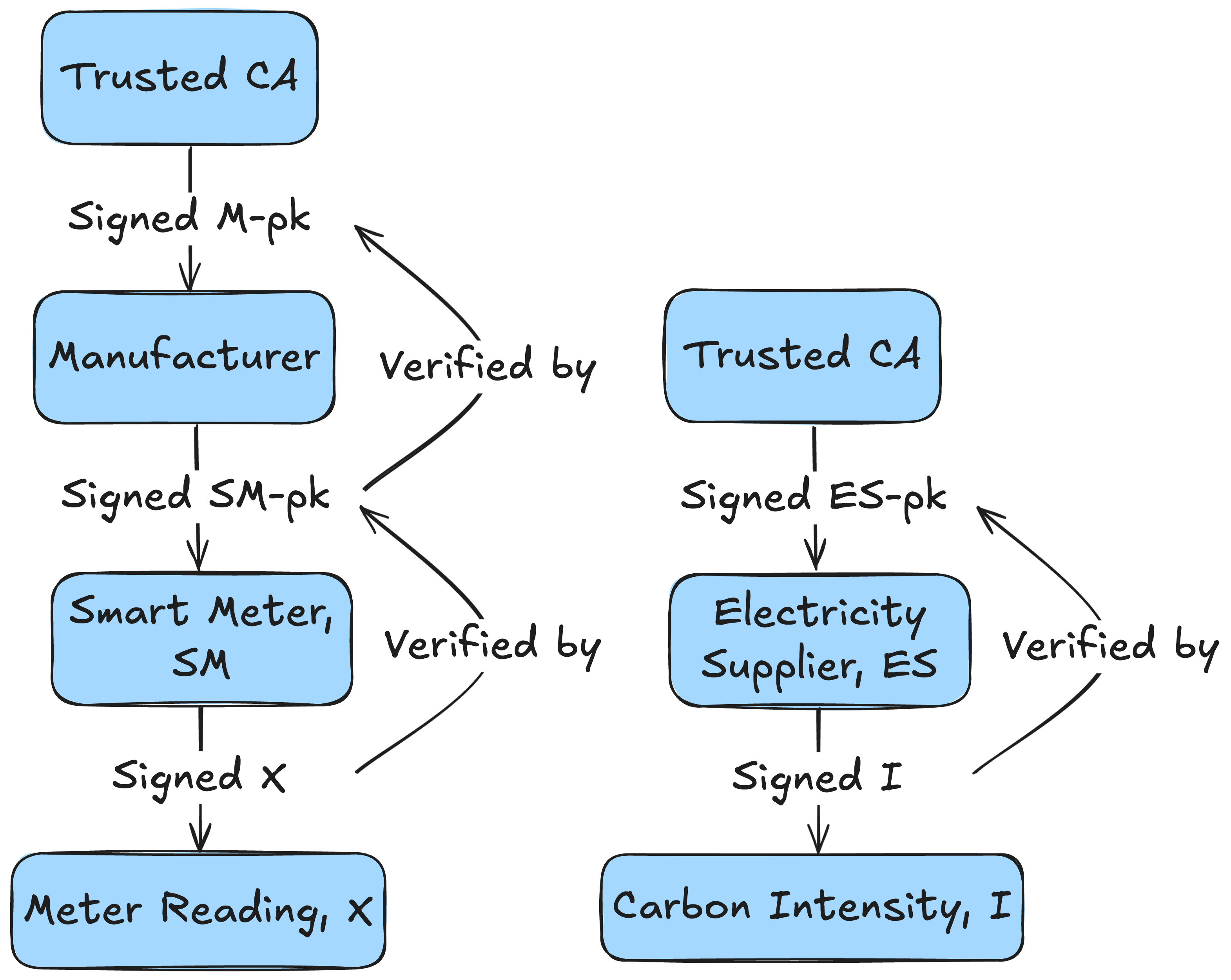}
    \caption{Signature chains for verifiable meter readings and carbon intensity}
    \label{fig:chained_signatures}
\end{figure*}

The same applies to the carbon intensity metric used in the accounting. We assume that the carbon intensity has been signed by the electricity supplier, and the public key of the electricity supplier has been signed by a trusted CA, which confirms that the key belongs to a genuine electricity supplier. Fig.~\ref{fig:chained_signatures} shows these certificate chains. In principle, electricity suppliers could use another zk-SNARK to prove that the carbon intensity figures have been calculated accurately, and the data centre operator's proof can recursively attest to the validity of the carbon intensity proof. We plan to explore recursive ZKPs in future work.

A zk-SNARK then allows a cloud provider to prove that there is a valid signature chain for every meter reading and every carbon intensity figure used in the emissions accounting, but without disclosing the value of those meter readings, the identity of the electricity supplier, or any other internals that might be sensitive. Only the public key of the CA needs to be disclosed, since it serves as the trust anchor for the computation.

A customer's share of total energy consumption is internal and sensitive data, therefore should not be disclosed. One way to check that each customer's share of usage is accurate is to apply the `Completeness Principle' defined in the Greenhouse Gas Protocol \cite{ghg_reporting_standards}, namely that the entire emissions of the data centre must be allocated to customers. That means that if we can prove that all customers' shares from the same data centre over the same period of time add up to 100\% of the power consumption, and that the verifier's share is one of them, then we have reasonable confidence that the input metric can be trusted. This extended proof is beyond the scope of this paper, but it is important to consider it as future work (see \S\ref{sec:caveats_and_future_work}).

%Link text to components on diagram
\subsection{Applying zk-SNARKs}
\begin{figure*}
    \centering
    \includegraphics[width=\linewidth]{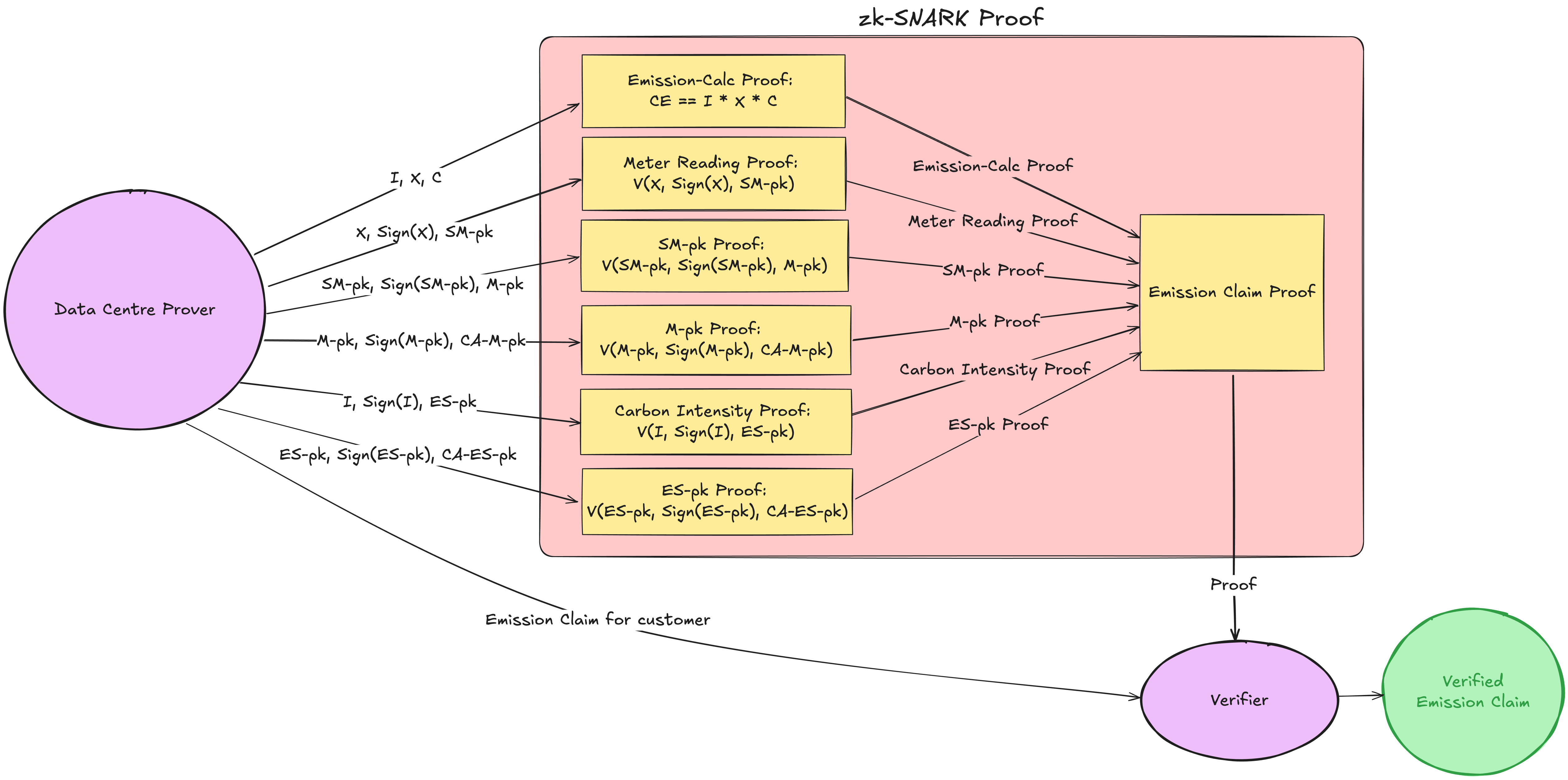}
    \caption{zk-SNARK proof system for private data protected carbon emissions claims. Private data: $I$: Carbon intensity, $X$: Data centre's total electricity consumption, $C$: Customer's share of the total electricity consumption, Sign(n): signature of n, $SM\textnormal{-}pk$: Smart meter's public key, $M\textnormal{-}pk$: Manufacturer's public key, $CA\textnormal{-}M\textnormal{-}pk$: CA's public key for verifying the signature of manufacturer's public key, $ES\textnormal{-}pk$: Electricity supplier's public key, $CA\textnormal{-}ES\textnormal{-}pk$: CA's public key for verifying the signature of electricity supplier's public key. Knowledge of verifier: $CE$: Emission claim}
    \label{fig:the_proof}
\end{figure*}

Now that we have defined what the witnesses are (input data) and what we need to prove (signature verification and carbon emission calculation), we can then apply a zk-SNARK in three stages, as illustrated in Fig.~\ref{fig:the_proof}.

\paragraph{Stage 1: Define computation} In traditional zk-SNARKs, we must define a circuit that encodes the accounting methodology used to calculate the emissions and any verification that needs to occur, such as checking the signatures in a certificate chain. The relationships among the input parameters are reduced to a set of polynomial equations, or \textit{constraints}. Domain-specific languages such as Circom \cite{belles2022circom} can be used to define the circuit in code, and the circuit is made public. This is illustrated in Fig.~\ref{fig:the_proof} as the centre box showing each proof generation on the left, and the public output, labelled ``Emission Claim Proof'' on the right.

Alternatively, there are also zero-knowledge virtual machines (zkVMs), such as SP1 \cite{sp1} and RISC Zero \cite{risc0}, that offer a more developer-friendly alternative to writing circuits by executing programs written in a conventional programming language such as Rust. These frameworks are widely used in the cryptocurrency community. We are exploring both the use of zkVMs and circuit-based techniques.

\paragraph{Stage 2: Generate proof} The prover proves that it knows a set of witnesses that satisfy all the constraints, and it can choose which of the witnesses to disclose and which to keep private. Public keys and signatures from smart meters, meter manufacturers, and electricity suppliers do not need to be disclosed to the verifier.

The prover in zk-SNARK requires more computational resources than the verifier. The proof can be constructed using one of the several zk-SNARK protocols, such as Groth16 \cite{groth2016size}, which is a pairing-based zk-SNARK widely used for its succinctness. Groth16 requires a trusted setup before a proof can be generated, but there are also alternative constructions that do not require any trusted setup. The prover then takes the proving key and witness to compute a proof.

In our example scenario, the cloud provider acts as the prover. They input the carbon emissions claim, carbon intensity for the electricity used by the data centre, the total power consumption, the signatures, public keys, and any other required data to create a proof that the witnesses satisfy the equations defined in the circuit.

An example of input, assuming that the public key of the smart meter manufacturer is signed by a CA using the EdDSA scheme \cite{rfc_eddsa}:

\begin{lstlisting}[breaklines=true, basicstyle=\footnotesize\ttfamily]
{
    "customer_emission":"xxxxx", (in kgCO2e)
    "carbon_intensity":"xxxxx",  (in kgCO2e/kWh)
    "total_consumption":"xxxxx", (in kWh)
    "customer_share":"xxxxx",    (in %)

    "ca_pk_x":"xxxxx",
    "ca_pk_y":"xxxxx",
    "manufacturer_pk_signature_r_x":"xxxxx",
    "manufacturer_pk_signature_r_y":"xxxxx",
    "manufacturer_pk_signature_s":"xxxxx",
    "hashed_manufacturer_pk":xxxxx
}
\end{lstlisting}

\paragraph{Stage 3: Verify} The verifier uses the proof and the knowledge they have to determine whether they can believe that the prover has knowledge of all input data, such that the private and public witnesses satisfy all the equations encoded in the circuit. That is,
    \begin{equation*}
        \exists \mathit{private\; witness}.\; C(\{\mathit{public, private}\}\, \mathit{witness}) = \mathit{True}
    \end{equation*}
    
The customer in our example, as verifier, can use the generated proof and the shared knowledge, namely the carbon emissions claim and the CA's public key, to verify the truthfulness of the claim, as shown at the bottom of Fig.~\ref{fig:the_proof}. The public witness sent to the verifier to the verification therefore contains only three numbers in our example:

\begin{lstlisting}[breaklines=true, basicstyle=\footnotesize\ttfamily]
{
    "customer_emission":"xxxxx", (in kgCO2e)
    "ca_pk_x":"xxxxx",
    "ca_pk_y":"xxxxx"
}
\end{lstlisting}

A more fully-fledged example would need to contain several more fields, such as the start and end timestamps of the reporting period, the identity of the data centre operator, and the ID of the customer for which the report has been generated.

\section{Caveats and Future Work} \label{sec:caveats_and_future_work}
The main challenge with the proposed approach is the accuracy and quality of the source data. A ZKP alone cannot guarantee everything. For example, if a cloud provider gets meter readings from multiple data centres, there is a chance that they might input the wrong ones or leave out the reading from some meters. If the meters are incorrectly connected, the readings would be wrong. The carbon intensity figures could also be applied inaccurately if they were meant for a different region. These problems would require manual checks outside of the ZKP. Separate validations to check that the data is consistent across multiple data centres would be useful, and it would make the validation stronger if the carbon-related data is linked with financial data, such that discrepancies would arise if incorrect data has been used for one side of the equation. For example, a ZKP can show that the total paid to all suppliers matches the number reported on the audited accounts and that these supplier transactions are also reflected in the emissions calculation.

With regard to smart meters themselves, previous work has looked at secure hardware in more detail. For example, Karakashev et al. looked at using secure hardware for trustworthy renewable energy certificates ~\cite{karakashev2020making}. Human auditing or random sampling verification can also be used to provide extra reassurance.

Another potential concern is the energy consumption allocation. There are shared resources, such as lighting, cooling, networking equipment and other ancillary energy consumers, that cannot be allocated directly to customers based on resource usage. One way to deal with this is to separate the emissions from customers' usage and shared or fixed resources, and to allocate shared resources across all customers proportionally to usage. For example, Google's carbon accounting methods first allocate emissions from electricity consumption per customer, and then augment the figures with proportional allocations of emissions arising from the non-electricity sources \cite{gcp_carbonfootprint}.

Companies could also make up customer shares, allocate shares dishonestly, or create fictitious customers and allocate emissions to them. Having a requirement for a public customer-based transparency log would help to make it difficult for companies to cheat. With a transparency log, customers could look up their entry to validate that they have been accounted for, and the data could be encrypted to protect confidentiality. 

There are still open questions about the proposed approach. First, for verifiable data exchange to work, it relies on standardised accounting schemes being adopted by the stakeholders, but how do we encourage companies to commit to these schemes in the first place? The good news is that there are emerging standards for exchanging emissions data between companies. WBCSD (as mentioned in \S\ref{sec:relate_work_accounting}) is leading the effort and has produced a set of standards for emissions data exchange. However, the data exchange methodology does not currently involve cryptographic verification. The proposal presented in this paper could be extended and integrated into their framework to achieve data quality and reliability. 

Second, the example used in this paper only considers one meter reading, one signature from each party on the chain, for one customer's carbon emissions claim. The proof system has to be able to scale when we consider many customers and frequent meter readings. Performing proofs involving many customers and many meter readings in a single zk-SNARK circuit quickly runs into scalability limitations; we are therefore investigating methods for breaking down such large proofs into many smaller proofs, which are more computationally feasible.

Third, as mentioned in \S\ref{sec:methodology}, the proposed approach needs to be extended to include a way to verify that all the customer's share of power consumption add up to 100\%. This can be achieved in multiple possible ways, but the details are beyond the scope of this paper.

%% file: 6_discussion.tex
It seems almost impossible to balance privacy and competitiveness needs with our urgent sustainability goals to reduce emissions, particularly in the ICT sector. The approach outlined in this paper is an example of how we can achieve privacy-preserving and trustworthy carbon emissions claims for data centres. To adopt the ZKP system, companies can apply carbon accounting alongside their financial accounting, which already needs to attribute the use of computing resources to individual customers for billing purposes. zk-SNARK proofs are small enough (a few kilobytes) to be bundled along with emissions reporting.

We believe that this proposal is a step forward for carbon emissions accounting to be public and explicit, making emissions tracking more accurate and comparable across companies. We are exploring how this could be exposed directly to end users via browser plugins, providing an end-to-end verifiable CO\textsubscript{2e} cost alongside conventional costs used by users to make their buying decisions (such as price, delivery time or distance to the service). Our overall aim is to drive demand-side pressure to reduce unnecessary emissions from data centre use by informing consumers about the environmental cost of their actions online.

The approach can also be used in other industries. For instance, in an automobile manufacturing supply chain, motor part manufacturers would not want to disclose where their factories are or some of the metrics regarding the production of their products. Equally, in a coffee making supply chain, coffee beans suppliers would not want to make public where they source the beans. Companies typically have a lot of confidential data that are part of the carbon emissions calculations, and it is important that they can disclose verifiable Scope 1, 2 and 3 emissions to earn trust and for regulatory reasons, and still be able to protect sensitive business data.